\documentclass[twocolumn,twoside,slac_two]{revtex4}
\usepackage{graphicx}
\usepackage{fancyhdr}
\begin{document}
%
%Title of paper
\title{Measurement of the $W +$ jets and $Z +$ jets Cross Section with the ATLAS detector}
%
% Repeat the \author .. \affiliation  etc. as needed
%
% \affiliation command applies to all authors since the last
% \affiliation command. The \affiliation command should follow the
% other information
\author{A.~Ahmad$^*$ on behalf of the ATLAS collaboration}
\affiliation{ %E-mail: Ashfaq.Ahmad@cern.ch\\
$^*$Department of Physics and Astronomy, State University of New York at Stony Brook, Stony Brook, NY 11794-3800, USA
}
\begin{abstract}
The study of $W$ or $Z$ boson with accompanying hadronic jets in final states is of high importance at hadron colliders both to understand Standard Model processes and to measure background to Beyond Standard Model physics searches. The presence of one or more jets in the final state increases the complexity for the reconstruction of leptons and of missing transverse energy. The ATLAS prospects for the cross section measurement of $W/Z$ + jets events at 14 TeV center of mass energy and integrated luminosity of $1~{\rm fb}^{-1}$ using fully simulated data are discussed. The statistical and systematic limitations are discussed in terms of probing perturbative QCD predictions and Monte Carlo generators.
\end{abstract}
%
%\maketitle must follow title, authors, abstract
\maketitle
\thispagestyle{fancy}
% body of paper here - Use proper section commands
% References should be done using the \cite, \ref, and \label commands
% Put \label in argument of \section for cross-referencing
%\section{\label{}}
%
%%%%%%%%%%%%%%%%%%%%%%%%%%%%%%%%%%
\section{Introduction}
The production of $W$ or $Z$ bosons in conjunction with jets is an interesting process in its own right as well as background to many Standard Model and Beyond Standard Model physics processes. Processes with $W/Z$ + jets in the final state are good candidates for testing Quantum Chromodynamics (QCD). The results of measurement can be compared directly with fixed-order predictions at leading order (LO) and next-to-leading (NLO) order in QCD and the gauge boson mass ensures that the interaction scale is sufficiently large for applying perturabtive calculations.

In this paper we present a feasibility study for the $W/Z$+jets cross section measurement at ATLAS~\cite{atlas} with $1~{\rm fb}^{-1}$ of fully-simulated Monte Carlo data, an integrated luminosity which can be accumulated within the first two years of running. The goal of the analysis is to test the performance of the lepton and jet triggering and reconstruction algorithms in high jet multiplicity events, to develop the necessary analysis techniques and to evaluate the statistical and systematic uncertainties. The primary end-results of the analysis are hadron level cross-sections. After a discussion of available tools in the next section, techniques used for data analysis are described and a comparison between measurement and theoretical predictions is made. Finally some conclusions are drawn at the end.
%%%%%%%%%%%%%%%%%%%%%%%%%%%%%%%%%%
\section{Monte Carlo Datasets}
The Monte Carlo data sets used in these studies are generated with ALPGEN~\cite{alpgen} interfaced with HERWIG~\cite{herwig} and using the leading order PDF set CTEQ6LL~\cite{cteq6ll}. The generation is done with a renormalization and factorization scale of $m^2_{V}+p^2_{T,V}$ ($V=W,Z$) and the datasets are obtained by merging samples of $W/Z$ + n partons (where n=0-5), weighted according to the expected cross-sections and using the MLM~\cite{alpgen} matching scheme.

%\indent 
For the comparison with the fixed order theoretical predictions, the merged data sets are normalized to the NLO cross-sections. Currently the NLO calculations can only be performed up to $W/Z$ + 2 jets, although there is ongoing work for the calculation of 3 jet final states. NLO is the first order at which the $W/Z$ + jets production cross sections have a reliable normalization~\cite{qcdprimer}. Cross-sections for  $W/Z$ + 0, 1 and 2 (3) jet final states can be conveniently calculated at LO and NLO (for $W/Z$ + 3 jets LO only) using the MCFM~\cite{mcfm} program, and it is from this program that we determine our reference cross-sections. The MCFM program cross-sections were generated using the CTEQ6.1 PDFs and a renormalization/factorization scale of $m^2_{V}+p^2_{T,V}$ where $V=W,Z$. PYTHIA~\cite{pythia} signal and background samples are generated with version 6.323 using the corresponding ATLAS underlying-event tune \cite{atlastune}.
%%%%%%%%%%%%%%%%%%%%%%%%%%%%%%%%%%
\section{Event Selection}
W and Z boson final states are selected through their leptonic decays. Events are required to have one or two isolated leptons with transverse momentum $p_{T}$ above a threshold (typically 15-25 GeV) and lie in the range $|\eta|<2.4$, excluding the barrel-to-endcap calorimeter crack region ($1.37<|\eta|<1.52$). The $Z$ boson selection requires two lepton candidates with an invariant mass close to that of the $Z$ boson mass peak, whereas the $W$ boson selection requires a single electron/muon and transverse missing energy above 25 GeV.

%\indent	
In this analysis jets are reconstructed using the standard ATLAS seeded-cone algorithm with a radius of $\Delta R = 0.4$, built from calorimeter towers and calibrated to the hadron level. The lepton and jet candidates must be separated by $\Delta R_{lj} >$ 0.4. It is required that the jet transverse momentum be larger than 40 GeV and that the jet be in the range $|\eta|<3.0$.
%%%%%%%%%%%%%%%%%%%%%%%%%%%%%%%%%%
%\begin{figure}[htb]
\begin{figure}[htb]
\centering
%%   Copyright (c) 2001 The Americ
\includegraphics[totalheight=60mm,width=80mm]{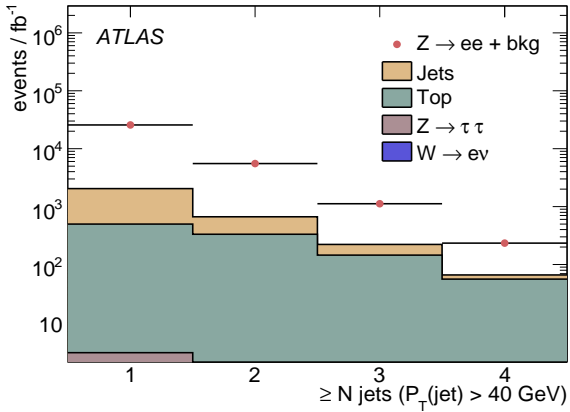}
\caption{Inclusive jet multiplicity distribution for $Z\rightarrow{ee}$ and background for jets with transverse momentum $p_{T} > 40$ GeV.} 
\label{bkgd:zeejets}
\end{figure}
%%%%%
%%%%%%%%%%%%%%%%%%%%%%%%
\subsection{Trigger and Reconstruction Efficiency}
In the $W/Z$ + jets analysis events are required to pass either double or single isolated lepton trigger. The trigger efficiencies are evaluated using the data driven tag-and-probe method on $Z\rightarrow{ll}$ events and compared to the Monte Carlo based method. Good agreement between the two methods is found. The trigger efficiencies are also studied as a function of the jet multiplicity and overall hadronic activity. In the presence of additional jets in the event, leptons are more boosted and the distance between leptons and jets become smaller. As a result, the trigger efficiency is found to decrease due to isolation requirement but the drop in efficiency is almost balanced by the larger acceptance resulting from the boost of leptons in multi-jet environment. %As a result, the larger acceptance resulting from the boost of leptons in multi-jet environment almost balances the drop in efficiency caused by trigger isolation requirements. 
The total reconstruction efficiency (offline+trigger) is therefore stable with respect to both the jet multiplicity and the transverse momentum of the leading jet.  
%%%%%%%%%%%%%%%%%%%%%%%%%%%%%%%%%%
\subsection{Background Estimation}
\begin{figure}[t]
\centering
%%   Copyright (c) 2001 The Americ
\includegraphics[totalheight=60mm,width=80mm]{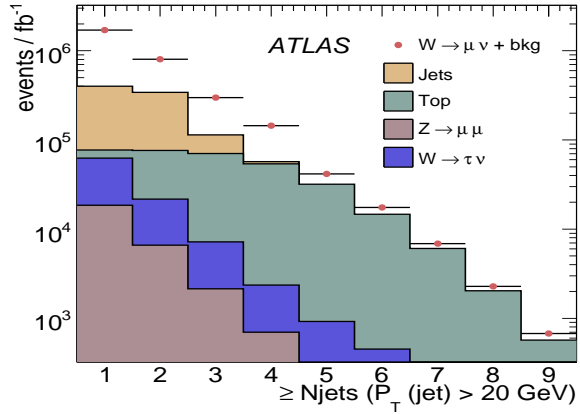}
\caption{Inclusive jet multiplicity distribution for  $W\rightarrow{\mu\nu}$ and background for jets with transverse momentum $p_{T} >$ 20 GeV.} 
\label{bkgd:wenujets}
\end{figure}

For the evaluation of backgrounds to the  $W/Z$ + jets different processes with real leptons ($t\overline{t}$, $W\rightarrow{l\nu}$, $Z\rightarrow{\tau^{+}\tau^{-}}$) and QCD multi-jet production are considered. Statistics of the multi-jet background sample are increased by applying a very loose electron selection and then re-weighting the events with the rejection from the final electron identification cuts. The dominant source of background is the production of QCD multi-jet final states where one or two jet fake a lepton candidate and in the case of $W$ boson, a significant mismeasurement of the jet energy results in large missing transverse energy. The background contamination is at the level of 5-15\% depending on the jet multiplicity as shown in Fig.~\ref{bkgd:zeejets} for $Z\rightarrow{e^{+}e^{-}}$ and in Fig.~\ref{bkgd:wenujets} for $W\rightarrow{\mu\nu}$. In both cases it varies with jet multiplicity. At low multiplicity QCD multi-jets are the dominant background where $t\overline{t}$ dominates at high multiplicity.
%%%%%
\section{Systematic Uncertainties}
\begin{figure}[htb]
\centering
%%   Copyright (c) 2001 The Americ
\includegraphics[totalheight=60mm,width=80mm]{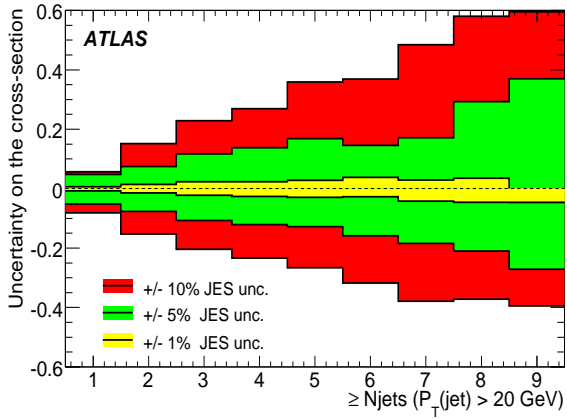}
\caption{Relative uncertainty on the $W\rightarrow{\mu\nu}$ differential cross-section for different levels of jet energy scale uncertainty.} 
\label{jes:uncertainty}
\end{figure}

The dominant source of experimental and theoretical uncertainties on all $W/Z$ + jets measurements are jet energy scale (JES) uncertainties and parton distribution functions (PDF) uncertainties, both of which affect signal efficiencies directly. After careful detector calibration has been applied using sophisticated techniques, some residual jet energy scale uncertainty still arises from the absolute energy correction, underlying event correction and energy outside the reconstructed jet cone. The JES uncertainty can shift the jet multiplicity spectrum and hence affect the  $W/Z$ + jets cross-section measurements. The effect of JES uncertainty on the cross-section measurement is shown in Fig.~\ref{jes:uncertainty}. It is clear that the effect due to JES uncertainty is significant. When the jet energy is miscalibrated by 1\% (which is the design goal of ATLAS) the experimental systematic uncertainity on the cross-section remains within 5\%, but they increase siginificantly with higher miscalibration.

\begin{figure}[t]
\centering
%%   Copyright (c) 2001 The Americ
\includegraphics[totalheight=60mm,width=80mm]{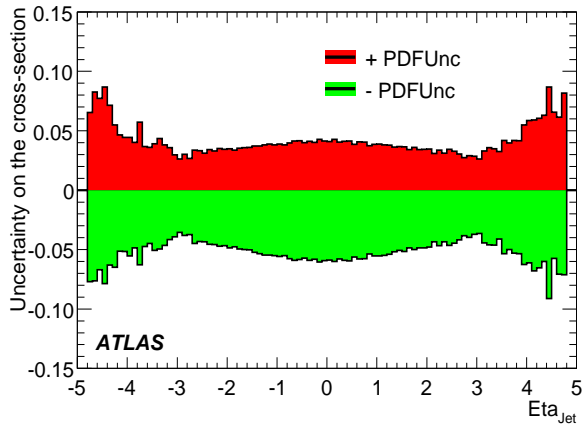}
\caption{Relative uncertainty on on the jet $\eta$ distribution from  $W\rightarrow{e\nu}$ events due to PDF uncertainties.} 
\label{pdf:uncertainty}
\end{figure}

%\indent	
The PDFs affect any cross-section calculation at the LHC, therefore their knowledge is vital for precise predictions of both Standard Model and beyond-standard-model processes. The PDF uncertainties are calculated by applying re-weighting technique. The ALPGEN datasets generated with the PDF set CTEQ6LL are re-weighted to the central value of CTEQ6M and the 40 error sets corresponding to the upper and lower uncertainty on each of the 20 eignvectors. Since jets are defined only within certain $\eta$ and $p_{T}$ ranges, a change in the PDFs modifies the jet distribution as shown in Fig.~\ref{pdf:uncertainty}.
%%
%%%  
%
\section{Unfolding of Detector Effects}
\begin{figure}[htb]
\centering
%%   Copyright (c) 2001 The Americ
\includegraphics[totalheight=60mm,width=80mm]{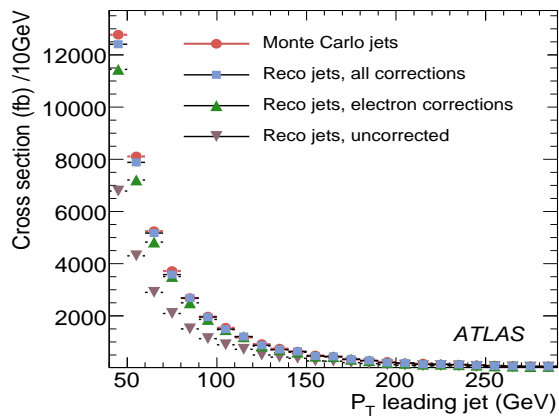}
\caption{Corrections to the jet distributions in $Z\rightarrow{ee}$ channel for electron reconstruction and trigger efficiency as well as corrections for the jet reconstruction efficiency and energy resolution.} 
\label{zee:correction}
\end{figure}

The reconstructed data have to be unfolded from the detector level to the particle (hadron) level, correcting for efficiency, resolution and non-linearities in electron and jet reconstruction. The unfolding is very important in order to compare data with theoretical calculation. The effects of applying correction for the electron reconstruction and trigger efficiency, followed by correction for the jet reconstruction efficiency and jet energy resolution are shown in Fig.~\ref{zee:correction}. The unfolded data is directly compared to jets formed from generator-level hadrons, after showering but before any detector interaction.
\section{Generator Comparison}
Predictions for the inclusive jet cross-sections from the Monte Carlo Generators PYTHIA and ALPGEN are compared to fixed-order LO and NLO calculations from MCFM partonic level generator. The NLO MCFM predictions for the $Z$+ 1 parton and $Z$+ 2 partons are of the order of 20 to 30\% greater than the LO predictions. The difference between PYTHIA and ALPGEN, and between both generators and MCFM amounts to 10-60\% depending on jet multiplicity. Both Monte Carlo generators predict a lower cross-section than the NLO MCFM caluclation for final states with more than one jet. PYTHIA predicts a larger $Z$ + 1 jets cross-section than ALPGEN but also predicts a lower average jet multiplicity. The difference between PYTHIA and ALPGEN depends very much on the minimum jet $p_{T}$ required by the selection. PYTHIA predicts larger cross-section than even NLO MCFM for low jet $p_{T}$. While the jet $p_{T}$ distribution predicted by ALPGEN agrees well with the NLO MCFM predictions, PYTHIA generates a clearly softer $p_{T}$ spectrum.
\section{Conclusions}
Final states containing $W/Z$ + jets will serve as one of the Standard Model benchmarks for physics analysis at the LHC while probing new QCD regimes.  We have considered cross-section measurements for theoretically well-defined quantities such as the inclusive $Z$ + jets cross section, and the jet transverse momentum for leading and next-to-leading jets. Preliminary studies have identified major sources of background and sources of systematic uncertainty in the cross-section measurements. Comparisons to theoretical predictions at the next-to-leading order level are facilitated by the experimental unfolding technique, which allows comparison at particle (hadron) level. The dominant systematic error is expected to be the uncertainty on the jet energy scale. It has been shown that for jet energy scale uncertainty of 5-10\%, the uncertainities on cross-section could be as large as 20-30\% which is at the same order as the typical differences expected between LO and NLO predictions or between predictions from PYTHIA, ALPGEN and MCFM. The ATLAS goal is to keep jet energy scale uncertainty within 1-2\%, hence uncertainties on cross-section would remain with 5\%.      
%   
%%%%%%%%%%%%%%%%%%%%%%%%%%%%%%%%%%
\bigskip % extra skip inserted
% Create the reference section using BibTeX:
%\bibliography{basename of .bib file}

%
\end{document}